# A Coarse-to-Fine Multi-stream Hybrid Deraining Network for Single Image Deraining

Yanyan Wei[1,2], Zhao Zhang[1,2], Haijun Zhang[3], Richang Hong[1,2] and Meng Wang[1,2]
[1] Key Laboratory of Knowledge Engineering with Big Data (Ministry of Education), Hefei University of Technology
[2] School of Computer Science and Information Engineering, Hefei University of Technology, Hefei, China
[3] Department of Computer Science, Harbin Institute of Technology (Shenzhen), Xili University Town, Shenzhen, China
e-mails: cszzhang@gmail.com, weiyanyan@hfut.mail.edu.cn

*Abstract—* Single image deraining task is still a very challenging task due to its ill-posed nature in reality. Recently, researchers have tried to fix this issue by training the CNN-based end-to-end models, but they still cannot extract the negative rain streaks from rainy images precisely, which usually leads to an over de-rained or under de-rained result. To handle this issue, this paper proposes a new coarse-to-fine single image deraining framework termed *Multi-stream Hybrid Deraining Network* (shortly, MH-DerainNet). To obtain the negative rain streaks during training process more accurately, we present a new module named dual path residual dense block, i.e., *Residual path* and *Dense path*. The Residual path is used to reuse common features from the previous layers while the Dense path can explore new features. In addition, to concatenate different scaled features, we also apply the idea of multi-stream with shortcuts between cascaded dual path residual dense block based streams. To obtain more distinct de-rained images, we combine the SSIM loss and perceptual loss to preserve the per-pixel similarity as well as preserving the global structures so that the deraining result is more accurate. Extensive experiments on both synthetic and real rainy images demonstrate that our MH-DerainNet can deliver significant improvements over several recent state-of-the-art methods.

*Keywords—* Single image deraining; dual path residual dense block; multi-stream hybrid deraining neural detwork

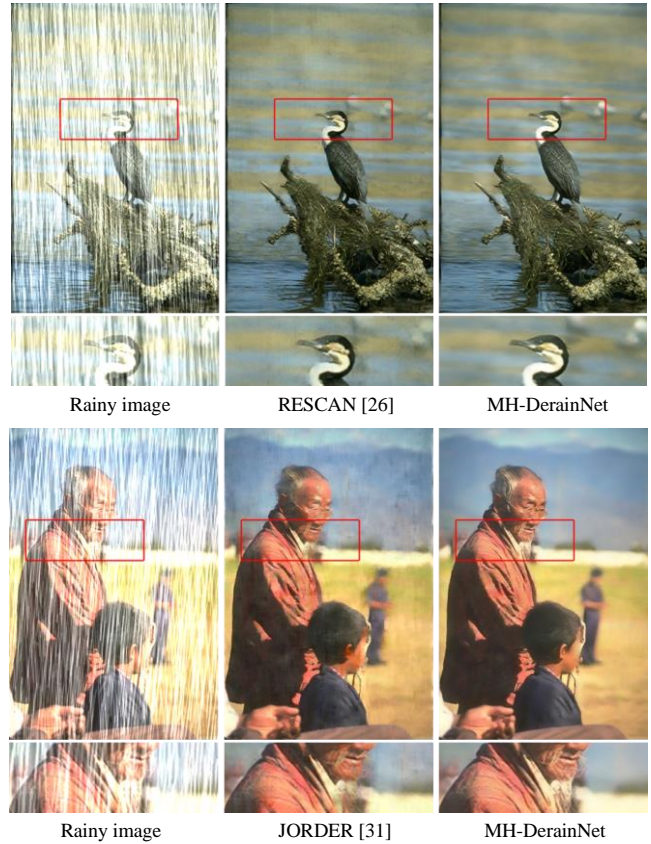

Fig. 1. Illustration of using MH-DerainNet to remove the heavy rain streaks from single images, comparing with two state-of-the-art models.

## I. INTRODUCTION

Removing rain streaks from images still remains an important and challenging topic in the outdoor tasks of computer vision and data mining [26, 27][30, 31][42, 43, 44], e.g., self-driving, drone-based video surveillance and real-time object recognition under severe rain weather conditions, etc. Since rain is one of the most common weather condition degrading the quality of images, but due to inappropriate expressions of images with rain streaks, the subsequent high-level tasks such as object detection [1], image recognition [8] and saliency detection [19] may be affected, so it is important to develop novel and effective models to remove rain streaks from images automatically.

Rainy images mainly contain the rain streaks and rain drops, which create a rain mask before true image. For heavy rains, they may cause a haze atmosphere due to light scattering and thus making images blurring and haziness. So, separating the rain mask from true image is an intuitional idea to solve this task. The rain-removal problem can be modeled as

$$X = R + B, \quad (1)$$

where *X* denotes a rainy image, which can be decomposed into a rain-streak component *R* and a clean background image *B*.

Different from the video (i.e., image sequence) based deraining methods [3, 4, 11, 22, 28, 32] which frequently use temporal content and inter-frame information in videos as an additional information, the single image based deraining models [2, 5, 6, 7, 9, 20, 21, 24, 25, 26, 27, 29, 31, 35] will be more challenging for lacking of aided information, e.g., temporal and inter-frame information. To address this ill-posed problem, various traditional methods [2, 9, 20, 24, 25, 35] have been proposed, such as adaptive nonlocal means filter to separate the rain streak from background [20], Gaussian mixture model (GMM) [24], sparse coding based models [2, 25] and low-rank representation-based methods [9, 35] to detect and remove the rain streaks. Although the traditional methods can handle this issue to some extent, they still cannot separate the rain streaks from the rainy image completely. Besides, due to the handcraft low-level representation produced by strong prior assumptions, the details in the output scenes tend to be overly smooth.

In recent years, due to the fact that deep learning has grown very fast and achieved significant success in both high-level and low-level vision tasks, thus many Convolutional Neural Network (CNN)-based methods have been proposed for single image deraining, such as [5, 6, 7, 26, 27, 29, 30, 31]. For thesemethods, they usually train an end-to-end deep network model to learn a mapping between rainy images and the corresponding ground truths that are regarded as no-rain images.

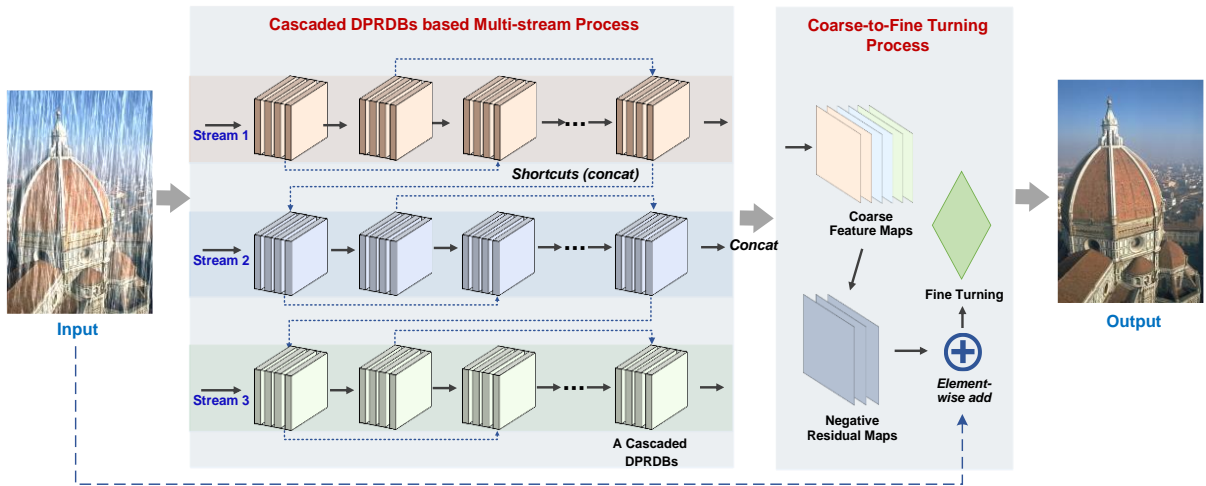

Fig. 2. An overview of the proposed MH-DerainNet framework. The input image is firstly to **cascaded dual path residual dense blocks (cascaded DPRDBs) based multi-stream process** where each stream has six cascaded DPRDBs. After that, the coarse rain streak feature maps extracted by multi-stream process will be concatenated together and sent to **coarse-to-fine turning process** to obtain negative residual maps. The negative residual maps will also be element-wise added with rainy image (input). Finally, a fine turning operation is conducted to get the precise derained image (output).

In [6], a novel large-scale dataset containing 14000 rainy/clean image pairs with 14 orientations and scales is synthesized, and then the ResNet model [12] is conducted as the parameter layers to learn a "neg-mapping" between clean and rainy images. In [31], a contextualized dilated network is used to jointly detect and remove rains from single images. Due to the complex ill-posed problem in Eqn. (1), the methods mentioned-above cannot extract exact rain streaks from rainy images, and often tend to over de-rain or under de-rain the image if the rain condition in the test image is not properly considered during training, as can be observed from Fig.1. The main reason maybe because these methods cannot obtain a precise mapping between input and output images for inadequate learning ability of their basic modules, especially in heavy rain conditions.

In this paper, we present a new coarse-to-fine multi-stream hybrid single image deraining network, which can potentially solve the above issues. The contributions are shown as:

(1) A new rain removal framework for single images, termed *Multi-stream Hybrid Deraining Network* (or shortly, MH-DerainNet), is technically proposed. Our MH-DerainNet can genially extract the rain streak features from rainy images by using a hybrid dual path residual dense block that can concatenate the rain streak features extracted from different streams to improve the performance. In addition, a hybrid dual path residual dense block based multi-stream connected network that uses different scale and shape information of rain streaks is built to obtain the negative map between input image and ground truth. We also consider the shortcut of original rainy image in the multi-steam process, which can improve the performance notably.

(2) To improve the representation of rain streaks, we propose a new hybrid block termed *dual path residual dense block* (DPRDB) as an elementary unit in the network to solve the single image deraining problem. The hybrid block is constructed by making the residual block and dense block be dual path, which can potentially make the hybrid block extract more precise negative residual than the other methods, especially in heavy rainy conditions due to the prominent capacity of detection heavy rain streaks.

(3) To preserve the per-pixel similarity as well as preserving global structures as much as possible, we define a hybrid loss based objective function using SSIM and perceptual losses. Based on the hybrid loss, the deraining results will be better than single SSIM loss or other complicated mixed loss based models that inevitably increase the difficulty of parameter tuning, while our hybrid loss is very simple with a ratio of 1:1 for SSIM and perceptual losses.

(4) Extensive simulations on several challenging synthetic datasets (i.e., Rain100H [31], Rain100L [31] and Rain12 [24]) and real images [6] demonstrate that our model can deliver state-of-the-art results, as can be seen from Fig.1.

This paper is outlined as follows. In Section II, we briefly introduce the related CNN based deraining methods and dual path network. Our MH-DerainNet is described in Section III. Experiment results on synthetic and real images are shown in Section IV. Finally, the paper is concluded in Section V.

## II. RELATED WORK

### A. Model-Based Single Image Deraining Methods

In real applications, removing rain streaks form a single image is an extremely challenging task for its ill-posed nature and lack of temporal information. To solve this issue, many deep learning-based models [5, 6, 7, 26, 27, 29, 30, 31] have been recently proposed, which aim to extracting the suitable low-level feature representation that are regarded as the rain streaks. These methods generally contain a main structure which plays an important role in their architectures and obtain precise low-level feature representation during the training process.

In [6], a deep detail network (DDN) that uses the residual block [12] is proposed to reduce the mapping range from input to output directly, which makes the learning process easier. In [29], a novel density-aware multi-stream densely connected convolutional neural network-based framework called

DID-MDN is presented for the joint rain density estimation and deraining, and the dense block [13] is the major constituent part. In [31], a multi-task deep learning architecture (JORDER) that learns the binary rain streak map, appearance of rain streaks and the clean background is proposed, which uses dilated convolution [10] module in a new contextualized dilated network to aggregate context information at multiple scales for learning the rain features. In [26], a novel deep network architecture based on deep convolutional and recurrent neural networks (RESCAN) which assigns different alpha-values to various rain streak layers according to the intensity and transparency by incorporating the squeeze-and-excitation block [14] is proposed. In [30], a novel residual-guide feature fusion network, i.e., ResGuideNet (RGN), is proposed for the single image deraining, which uses recursive convolution [16] to build each block and uses supervision to the intermediate results progressively to predict high-quality reconstruction.

In general, the basic modules of existing deep networks usually use the residual block [12], dense block [13][41], squeeze-and-excitation block [14], or novel convolutional operations such as the dilated convolution [10], squeeze-and-excitation. These modules have been proved to be not only effective for the high-level tasks, such as object detection [1], image recognition [8], but also effective for the low-level tasks, e.g., single image deraining [5, 6, 7, 26, 27, 29, 30, 31], image inpainting [23], image colorization [38] and image super resolution [15, 17]. While in this paper, we present a new dual path residual dense block as the basic modules of our network.

### B. Dual Path Neural Network

In this subsection, we will briefly review the dual path neural network, which motivates us to present the dual path residual dense block. In [18], a simple modularized and highly efficient deep model called Dual Path Network (DPN) is proposed for image classification, where it has presented a novel form of connection paths internally. By revealing the equivalence of the state-of-the-art Residual Network (ResNet) [12] and the Densely Convolutional Network (DenseNet) [13] in terms of mathematic formulation, it can be found that ResNet enables feature re-usage while DenseNet enables new feature exploration, which are all important for learning good representations. Extensive experiments on high-level tasks, e.g., image classification, demonstrate that this network has many advantages, such as high accuracy, small model size, low computational cost and low GPU memory consumption. Thus this network is very useful for the research on real-word applications.

Thus, inspired by this article, to inherit the benefits from both residual block and dense block for low-level feature representation, we propose to construct a hybrid block called dual path residual dense block (DPRDB) to handle low-level task, i.e., single image deraining. Based on the advantages of feature re-usage by residual block and new features exploration by dense block, our proposed DPRDB can obtain a precise negative rain residual map to mapping a rainy image, which will be verified by extensive simulations.

### III. COARSE-TO-FINE MULTI-STREAM HYBRID DERAINING NETWORK (MH-DERAINNET)

In this section, we describe the proposed MH-DerainNet in detail. The whole framework of MH-DerainNet is shown in Fig.2. As can be seen, there are two primary processes in our MH-DerainNet, i.e., *cascaded dual path residual dense block (cascaded DPRDBs) based multi-stream process* and *coarse-to-fine turning process*. The multi-stream process is mainly used to extract the coarse rain streak feature maps from a rainy image while the coarse-to-fine turning process is mainly used to produce a corresponding negative residual map through these coarse rain streak feature maps and obtain the final derained image. Next, we will detail the two processes. Besides, we will also introduce the basic DPRDB module and the multi-stream containing the cascaded DPRDBs. The hybrid loss based objective function will also be presented.

### A. Cascaded DPRDBs based Multi-stream Process

The multi-stream process in our MH-DerainNet contains three independent streams. A single stream has 6 cascaded DPRDBs, where each DPRBD uses a design of dual path hybrid block. Hence, the dual path residual dense block plays the core role in our cascaded DPRDBs based multi-stream process. Due to the ability of utilizing advantages of both residual block and dense block, DPRDB can reuse previous features and exploring new features, which will lead to brilliant learning capacity. So, cascaded DPRDB based multi-stream process can potentially extract more precise feature maps than existing process. Next, we show the DPRDB, cascaded DPRDBs and the cascaded DPRDBs based multi-stream process, respectively.

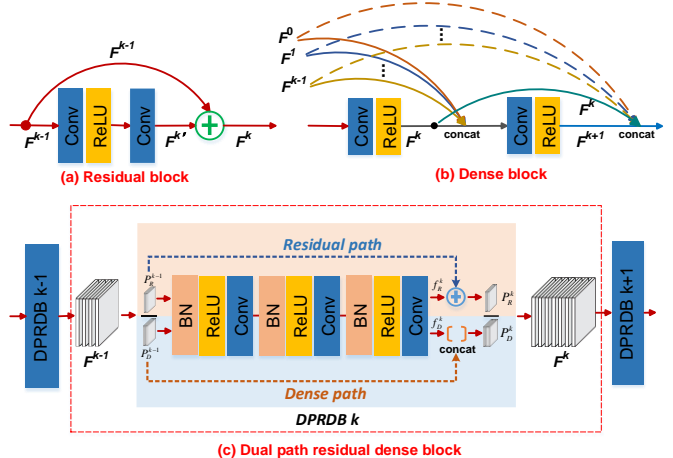

Fig. 3. Comparison of prior residual block of JORDER [31] in (a), prior dense block of DID-MDN [29] in (b), and our dual path residual dense block (DPRDB) in (c).

**Dual Path Residual Dense Block (DPRDB).** A DPRDB structure is shown in Fig. 3(c). It is clear that it contains two divided paths in DPRDB, i.e., *Residual path* and *Dense path*, respectively. The residual path containing the residual block is mainly to reuse common features from previous layers while the dense path containing the dense block can explore new features. By flowing through two paths, the feature maps will disperse and fuse in each DPRDB, so the network can potentially obtain more powerful feature representation ability. As a comparison, we also illustrate the residual block and dense block in Fig.3(a) and Fig.3(b), respectively. Note that the residual block uses a residual add operation between the original and conducted feature maps, so it can reuse previous feature maps and overcome drawback of gradient vanishing of deep layers. The dense block can sharply decrease the counts of network parameters and also can alleviates the gradient vanishing problem. Although some singe image deraining approaches use the residual block or dense block as main component, they only use a single path to obtain the feature maps in training process, so their deraining results may be inaccurate in reality, compared with our proposed dual path residual dense block.

In what follows, we will introduce the formulation of the DPRDB. Let $F^{k-1}$ and $F^k$ be the feature maps extracted from *(k-1)*-th and *k*-th DPRDB, respectively. Firstly, $F^{k-1}$ will be sent to *k*-th DPRDB for dual path operation. Specifically, in the *k*-th DPRDB, $F^{k-1}$ will be de-concatenated into two parts of feature maps for the subsequent use:

$$F^{k-1} = \left[ P_R^{k-1}, P_D^{k-1} \right], \qquad (2)$$

where $P_R^{k-1}$ and $P_D^{k-1}$ are the de-concatenate parts of $F^{k-1}$, $[\cdot]$ denotes the concatenate operation. Though we de-concatenate $F^{k-1}$ into two parts, we still use the original feature maps $F^{k-1}$ to obtain new feature maps. Then, we de-concatenate $f^k$ into two path components, which are shown as follows:

$$f^k = H_{BRC}^k \left( F^{k-1} \right), \qquad (3)$$

$$f^k = \left[ f_R^k, f_D^k \right], \qquad (4)$$

where $H_{BRC}^k(\cdot)$ denotes three BN-ReLU-Conv operations with kernel sizes of 1×1, 3×3 and 1×1, respectively. $f^k$ denotes the feature maps obtained after above operations. $f_R^k, f_D^k$ denote the feature maps that are de-concatenated from $f^k$, which will be used in the residual path and dense path, respectively. For the residual path, we add $P_R^{k-1}$ to $f_R^k$ in an element-wise way, and for dense path, we concatenate $P_D^{k-1}$ and $f_D^k$ together:

$$P_R^k = P_R^{k-1} + f_R^k, \qquad (5)$$

$$P_D^k = \left[ P_D^{k-1}, f_D^k \right], \qquad (6)$$

where $P_R^k$ denotes new feature maps obtained in residual path and + denotes element-wise add operation. Due to the nature of residual element-wise add operation, $P_R^k$, $P_R^{k-1}$ and $f_R^k$ will have the same dimension. $P_D^k$ denotes the new feature maps in dense path. Due to the nature of dense concatenate operation, $P_D^k$ will have an increased dimension for $f_D^k$ than $P_D^{k-1}$, and we define $k_D$ as the increase value of dimension in dense path. Finally, we can easily concatenate $P_R^k$ and $P_D^k$ together to obtain the new feature maps for the *k*-th DPRDB:

$$F^k = \left[ P_R^k, P_D^k \right]. \qquad (7)$$

By comparing with $F^{k-1}$ in the *(k-1)*-th DPRDB, $F^k$ has an increased dimension of $k_D$. After conducting these operations in the *k*-th DPRDB, $F^k$ can be similarly sent to the *(k+1)*-th DPRDB to obtain the new feature maps.

**Cascaded DPRDBs.** Based on the DPRDB defined above, we can easily connect several DPRDBs together to construct BLOCK structure called cascaded DPRDBs as illustrated in a dotted rectangle of Fig.4. With the help of cascaded DPRDBs, we can use the advantages of dual path residual dense block design. Specifically, the deeper the dual path goes, the more complex the feature maps will be and the more precise the representation will be. In this paper, we can connect 6 DPRDBs together to construct a cascade BLOCK in the simulations. Due to the extension of $k_D$ dimension in every dense path operation of DPRDB, the dimension of feature maps will grow fast and will cause a flood tide of parameters, so we use a transition layer as a tail to reduce dimension of feature maps, the operation of the transition layer is formulated as follows:

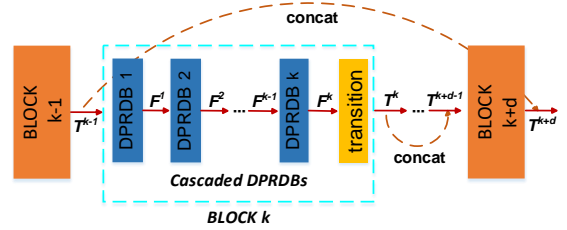

Fig. 4. The block structure of the cascaded DPRDBs (in dotted rectangle)

$$T^k = trans\left( F_k^k \right), \qquad (8)$$

where $F_k^k$ denotes the feature maps conducted from the *k*-th DPRDB of the *k*-th cascaded DPRDBs, and $trans(\cdot)$ denotes the BN-ReLU-Conv operation with kernel size being 1. $T^k$ is the feature maps extracted from the *k*-th cascaded DPRDBs.

In our network, since each single stream contains 6 cascaded DPRDBs, whose structures can be described as follows:

***BLOCK (6,1) – BLOCK (6, 1) – BLOCK (6, 1) –
BLOCK (6, 1) – BLOCK (6, 1) – BLOCK (6, 1),***

where ***BLOCK (6, 1)*** denotes a cascaded DPRDBs that has 3 DPRDBs and 1 transition layer in it. The kernel sizes of the convolution operation are 3×3, 5×5, 7×7 in three streams, respectively, and we use full padding for enabling the feature maps to preserve the same size.

To better reuse the previous feature maps, we also use the shortcuts between cascaded DPRDBs, i.e., *2*-th and *4*-th, *1*-th and *5-th* cascaded DPRDBs. So, $T^{k+d}$ can be defined as

$$T^{k+d} = H_{block}^{k+d}\left( \left[ T^k, T^{k+d-1} \right] \right), \qquad (9)$$

where $T^k, T^{k+d-1}$ and $T^{k+d}$ denote feature maps extracted from *k*-th, *(k+d-1)*-th and *(k+d)*-th cascaded DPRDBs, respectively. $H_{block}^{k+d}(\cdot)$ denotes the BN-ReLU-Conv operation on the *(k+d)*-th cascaded DPRDBs.

**Cascaded DPRDBs based multi-stream design.** Due to various orientation and density of rain streaks in rainy image, multi-scale features extracted from different kernel sizes of multi-stream can capture different information in rainy images. In fact, some recent deraining approaches have proved that using the multi-stream design can improve the deraining performance [29][31]. Thus, we also apply the idea of multi-stream design, but note that we have used a cascaded DPRDBs based multi-stream process, where each stream contains 6 cascaded DPRDBs. Based on the powerful learning ability of DPRDB, our new multi-stream can extract different feature maps with different receptive fields (i.e., 3×3, 5×5 and 7×7). To improve the information flow among different streams and to leverage features from different scales, we create shortcuts among features from different streams to strengthen feature aggregation and obtain better convergence. After obtaining features from multi-streams, we concatenate them together and sent it into coarse-to-fine turning process to get the final de-rained image.

### B. Coarse-to-fine Tuning Process

**Negative residual maps and fine-turning.** After extracting coarse rain streak feature maps from the cascaded DPRDBs based multi-stream process, we concatenate them together and feed it into a convolutional operation to get negative residual maps, then we element-wise add it with the input image

(rainy image) to get the de-rained image. For a better display quality, a fine-turning operation is used to refine the coarse de-rained image to the final fine exquisite de-rained image:

$$NR = tanh\left(H_{RES}\left(\left[S_1, S_2, S_3\right]\right)\right), \quad (10)$$

$$B = G(X + NR), \quad (11)$$

where $S_1, S_2$ and $S_3$ are the feature maps extracted from each stream respectively, $H_{RES}(\cdot)$ is a Conv operation with kernel size of 3 x 3, $tanh(\cdot)$ is a tanh operation, $NR$ and $X$ denote the negative residual of rainy image and rainy image respectively, + denotes the element-wise add operation, $G(\cdot)$ is a ReLU-Conv-ReLU-Conv operation with kernel size of 7x 7 and 3 x 3 respectively, and $B$ is the final no-rain image.

*C. The Objective Function*

We describe the objective function of our deraining network. Specifically, due to the success of the perceptual loss [34] in many low-level tasks [5, 7, 15, 17, 23, 26, 38], and the goal of outputting derained image to approximates its corresponding ground truth image when using the SSIM loss [33] as the evaluation metric, we creatively combine the perceptual loss with SSIM loss together, which can preserve the per-pixel similarity as well as preserving the global structures. Thus, the overall hybrid loss function for MH-DerainNet is

$$L = L_{ssim} + \lambda_P L_P, \quad (12)$$

where $L$ is the objective function of MH-DerainNet, $L_{ssim}$ is the SSIM loss which is used for measuring the similarity of two images, and $L_p$ is perceptual loss for minimizing the perceptual difference between de-rained image and the ground truth image. $\lambda_p$ is a tradeoff parameter. In the ablation study, we will show that this hybrid loss can obtain the best results than other forms of losses. Next, we will briefly introduce the formulations of the perceptual loss and SSIM loss.

***Perceptual loss:*** The perceptual loss is defined as follows:

$$L_p = \frac{1}{CWH}\sum_{c=1}^{C}\sum_{w=1}^{W}\sum_{h=1}^{H}\left\|F(\hat{x})^{c,w,h} - F(x)^{c,w,h}\right\|_2^2, \quad (13)$$

where $\hat{x}$ is the de-rained image by our MH-DerainNet and $x$ is corresponding ground truth image of rainy image. $F(\cdot)$ represents a non-linear CNN transformation. In our method, we use layer ReLU2_2 of the VGG-16 model [8], and we assume that the features are of size $w \times h$ with $c$ channels.

***SSIM loss:*** The SSIM loss is defined as the following:

$$SSIM(x, y) = \frac{(2\mu_x\mu_y + c_1)(2\sigma_{xy} + c_2)}{(\mu_x^2 + \mu_y^2 + c_1)(\sigma_x^2 + \sigma_y^2 + c_2)}, \quad (14)$$

$$L_{ssim} = -SSIM(\hat{x}, x), \quad (15)$$

where $SSIM(\cdot)$ is the SSIM function to calculate the similarity between two images, $x$ and $\hat{x}$ are also the original and de-rained images. Note that we aim at maximizing the SSIM value as much as possible, so $L_{ssim}$ is a negative function.

## IV. EXPERIMENTS AND RESULTS

In this section, we evaluate MH-DerainNet for single image deraining, along with illustrating the comparison results with several widely-used models, including one traditional method (i.e., GMM [24]), four CNN-based deep network models, i.e., DDN [6], RESCAN [26], DID-MDN [29] and JORDER [31], and one lightweight method (i.e., RGN [30]). In this study, three synthetic rain image datasets, i.e., Rain100H [31], Rain100L [31] and Rain12 [24], and real rainy images [6], are used to evaluate the deraining performance of each model. In addition, we also prepare the ablation study to discuss the important of components of our MH-DerainNet framework.

In this study, we perform the training and test using Pytorch platform [36] in Python environment on a NVIDIA GeForce GTX 1080i GPU with 12GB memory. We use Adam [37] as optimizer with an initial learning rate 0.001, which is decayed by multiplying 0.2 in every 30 epochs. For extending the training dataset, we create 15 patches in a 100 x 100 size per image. The size of training batch is 14 and 100 epochs will be trained, and the tradeoff parameter $\lambda_p$=1. For fair comparison, we use Peak Signal to Noise Ratio (PSNR) [39], Structural Similarity Index (SSIM) [33] as the quantitative metrics to evaluate the deraining performance of each model.

*A. Deraining Results on Synthetic Rainy Images*

We first evaluate each deraining model on three popular synthetic rain image datasets, i.e., Rain100H [31], Rain100L [31] and Rain12 [24]. Rain100H has five streak directions, which contains 1,800 rainy images for training and 100 rainy images for testing. Rain100L is a synthesized dataset with only one type of rain streak, which contains 200 rainy images for training and 100 rainy images for testing. Rain12 has one type of rain streak, which only contains 12 rainy images for testing and without rainy images for training, so we will use the model trained on Rain100L for testing. In Fig.5, we show some rain image samples from these synthetic rain image datasets.

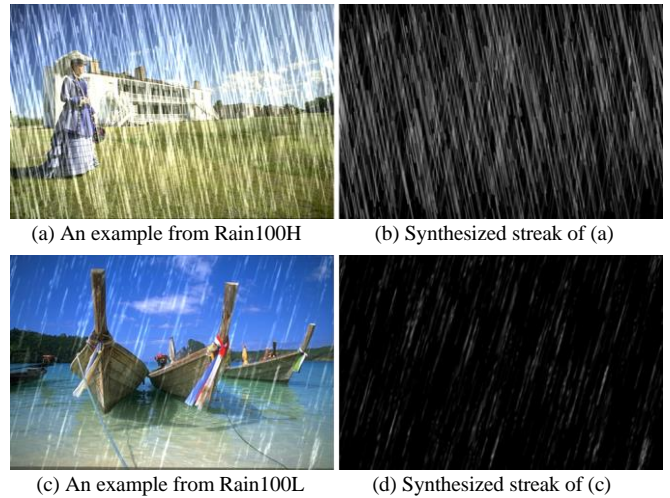

(a) An example from Rain100H    (b) Synthesized streak of (a)

(c) An example from Rain100L    (d) Synthesized streak of (c)

Fig. 5. The examples of synthesized rain streaks and rainy images.

The quantitative image deraining results in terms of PSNR and SSIM are shown in TABLE I. We can find that our MH-DerainNet achieves the best performance among all the testing methods, especially on the Rain100H dataset, which may be because MH-DerainNet uses the DPRDB to absorb more complicated rain streaks information than other deraining methods. The recently-proposed RESCAN also performs well by delivering highly-competitive results to our DPRDB.

Some visualized examples of derained images based on the Rain100H are shown in Figs.6 and 7, and those on Rain100L are shown in Fig. 8. In addition to evaluating the results by visual observation, we also provide the quantitative evaluation results. Specifically, we quantify the deraining results by computing the PSNR and SSIM values based on the derained image by each method with its corresponding ground true image. From the visual deraining results on Rain100H and Rain100L, we can find that: (1) MH-DerainNet can achieve the best results with the least artificial elements and without any slender rain streaks; (2) JORDER tends to generate artificial elements and blurring images as a result. Recent RESCAN generates few artificial elements than JORDER, but it still remains lots of slender rain streaks which makes the deraining results unreal. These results can once again demonstrate the effectiveness of our proposed coarse-to-fine multi-stream hybrid deraining network; (3) Based on the quantitative results, we find that our MH-DerainNet still outperforms the other algorithms, which keeps consistent with the visual observation results.

### B. Deraining Results on Real-world Rainy Images

In this study, we also evaluate the image deraining effects of each algorithm by using some real-world rainy images [6] that are widely used in single image deraining methods. Since there are no corresponding ground truth images to these real-world images, the de-rained images are only evaluated by human visual perception. The image deraining results are shown in Fig.9. We can find that DDN has the worst performance since it remains lots of rain streaks. JORDER [31] can also obtain better result, but it still has two potential drawbacks: (1) it changes the brightness of the original image which makes some pixels to become dark; (2) it tends to blur the background of the original image. RESCAN [26] remains lots of slender rain streaks. In contrast, our MH-DerainNet delivers the best performance.

TABLE I. QUANTITATIVE EXPERIMENTAL EVALUATION RESULTS ON DIFFERENT TESTSETS.

| Dataset | Rainy images | GMM [24] | RGN [30] | DDN [6] | JORDER [31] | DID-MDN [29] | RESCAN [26] | MH-DerainNet |
|---|---|---|---|---|---|---|---|---|
| **Rain100H** | 13.56/0.379 | 15.05/0.425 | 25.25/0.841 | 21.92/0.764 | 26.54/0.835 | 17.39/0.612 | 28.88/0.866 | **30.76/0.915** |
| **Rain100L** | 26.90/0.838 | 28.66/0.865 | 33.16/0.963 | 32.16/0.936 | **36.61/0.974** | 25.70/0.858 | -- | **36.56/0.975** |
| **Rain12** | 30.14/0.855 | 32.02/0.910 | 29.45/0.938 | 31.78/0.900 | 33.92/0.953 | 29.43/0.904 | -- | **36.12/0.958** |

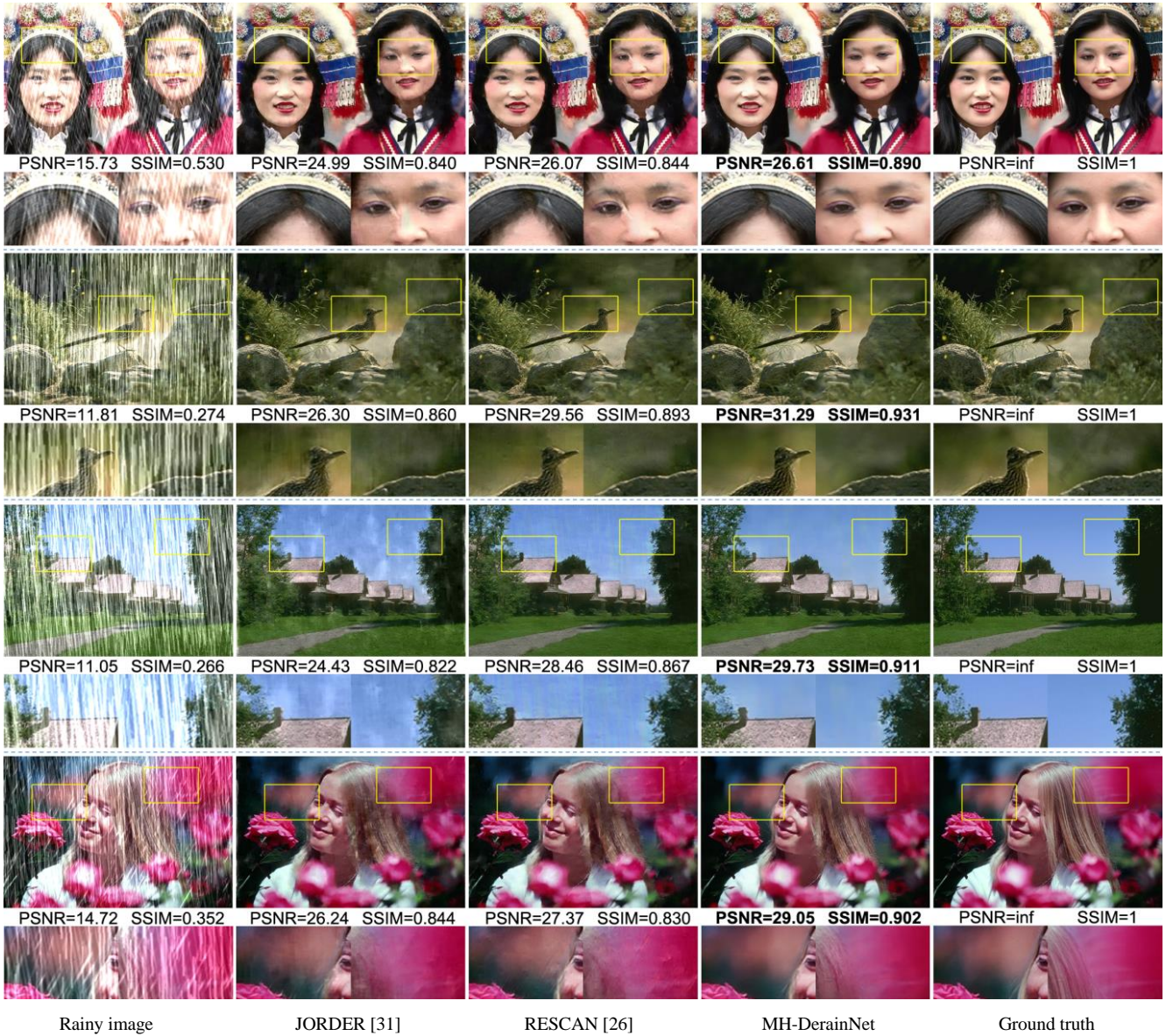

Rainy image　　　JORDER [31]　　　RESCAN [26]　　　MH-DerainNet　　　Ground truth

Fig. 6. Visual quality comparison of deraining by each deraining network models based on four images from Rain100H [31].

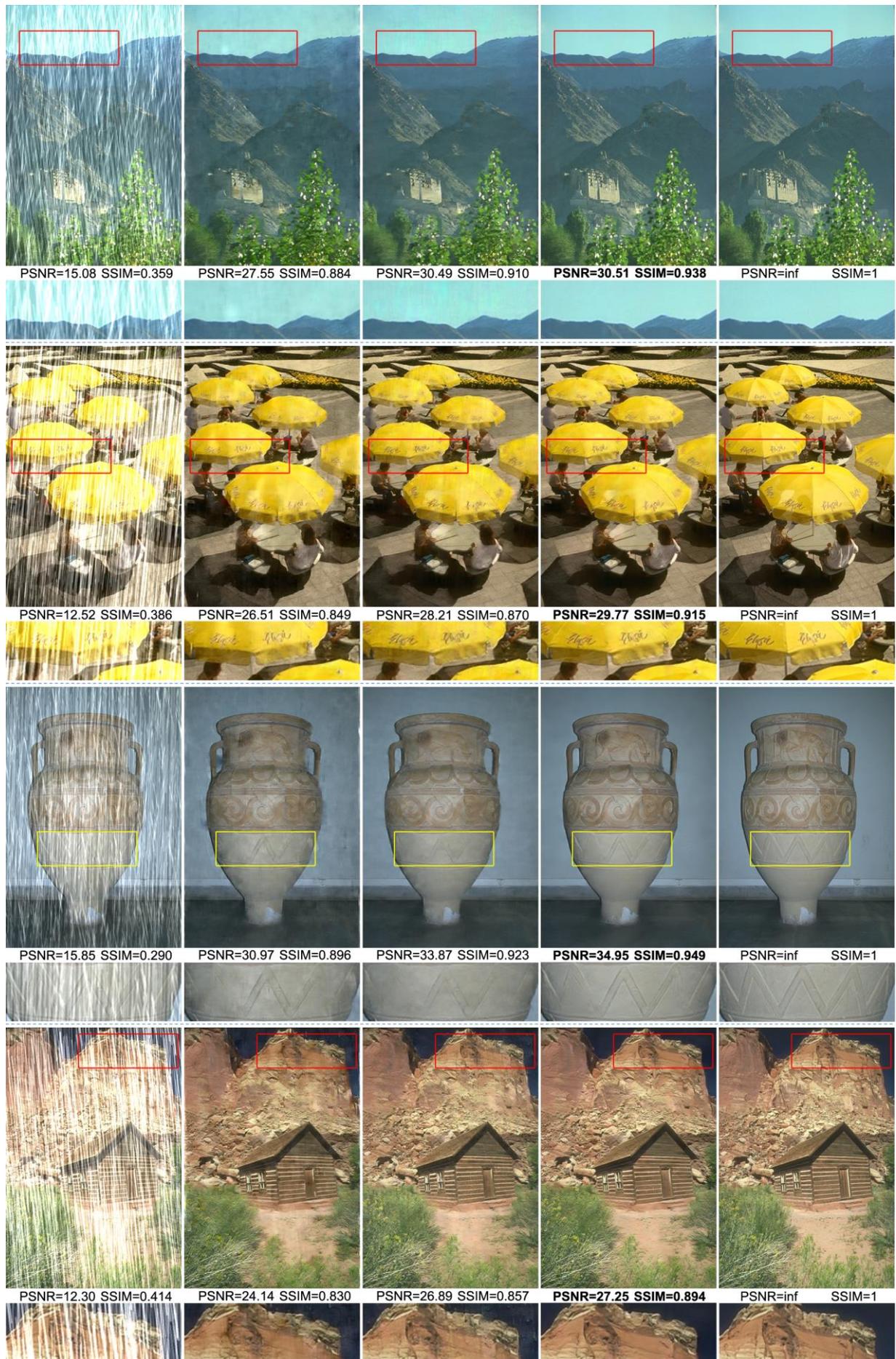

Fig. 7. Visual quality comparison of deraining by each deraining network models based on three images from Rain100H [31].

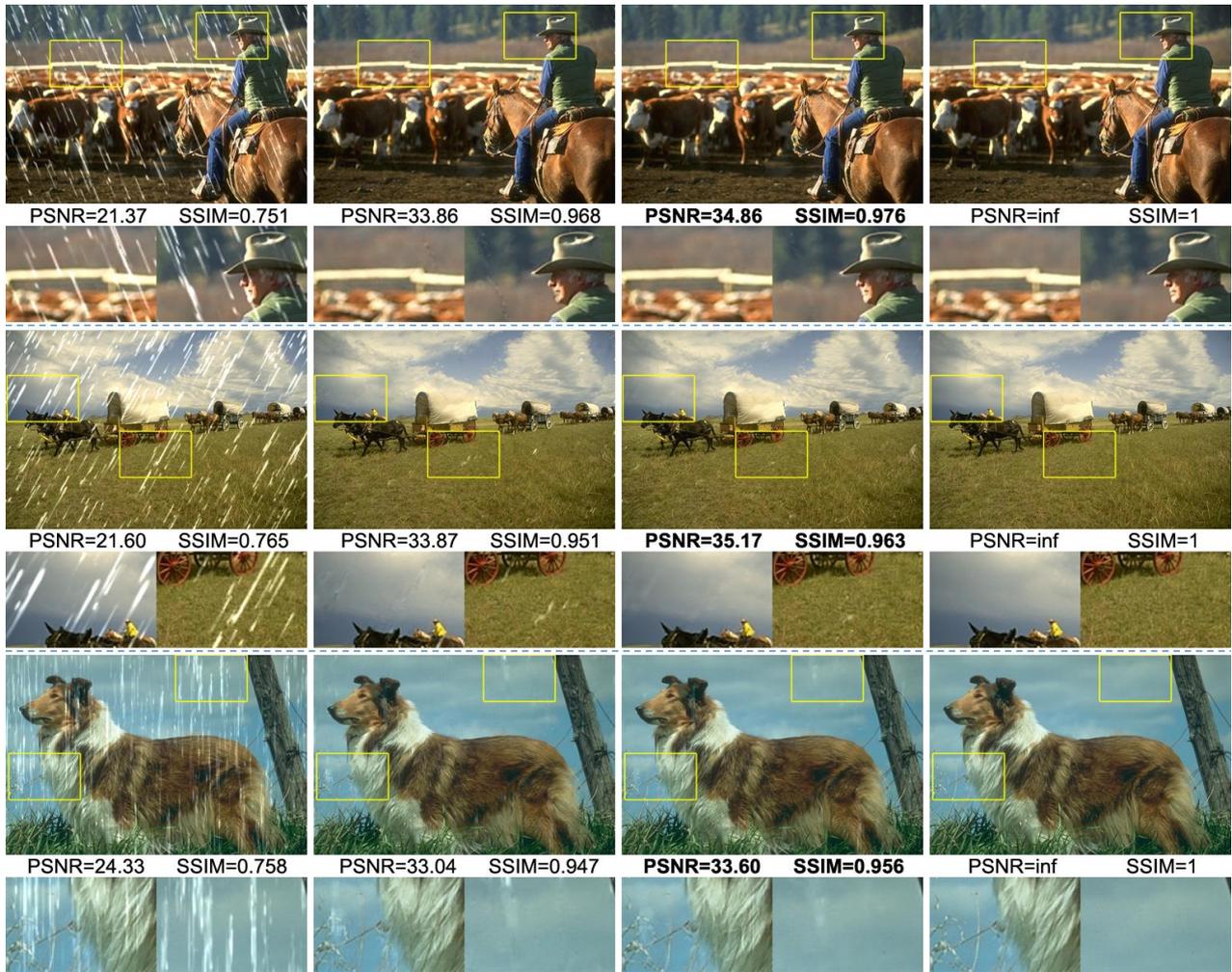

Fig. 8. Visual quality comparison of deraining by each deraining network model based on two images from Rain100L [31].

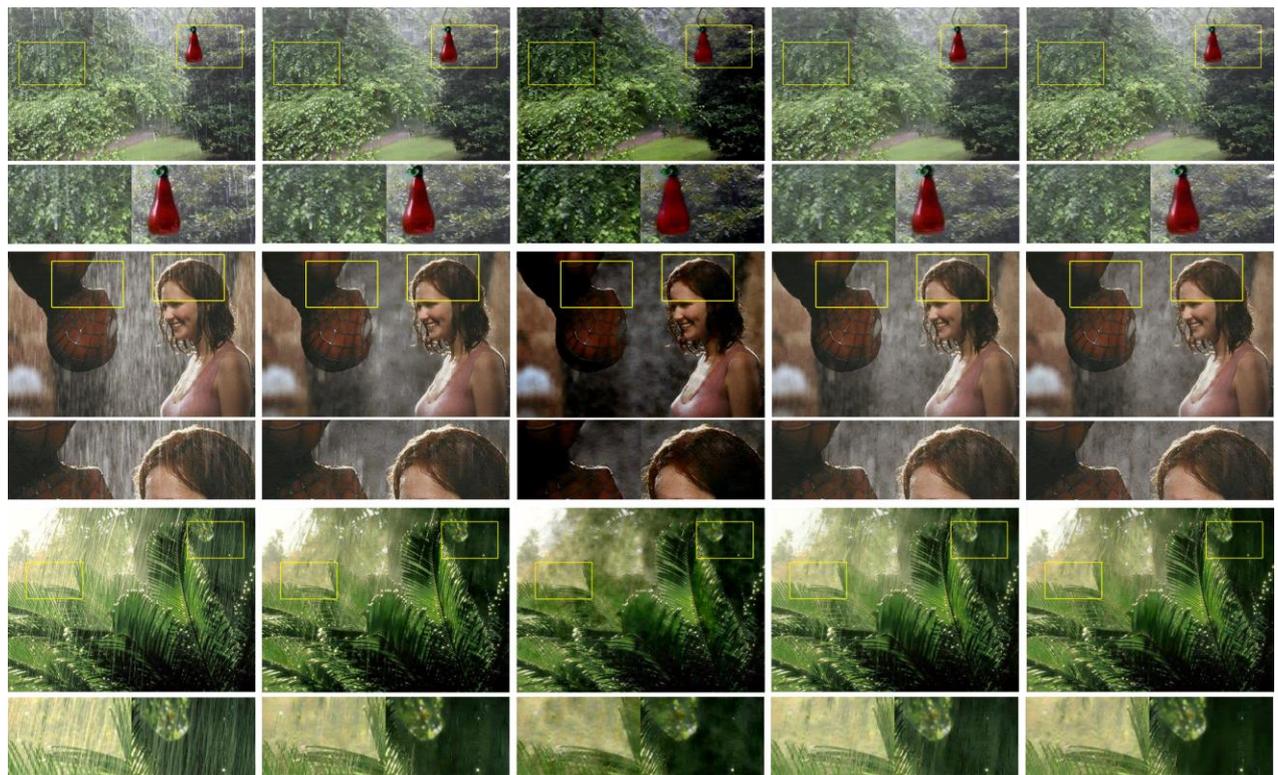

Fig. 9. Visual quality comparison of deraining by each deraining network model based on three real rainy images [6].

TABLE II. DERAINING COMPARISION OF MH-DERAINNET WITH DIFFERENT LOSSES.

| Dataset | $L_{mse}$ | $L_{ssim}$ | $L_{mse} + L_p$ | $L_{ssim} + L_p$ |
|---|---|---|---|---|
| Rain100H | 29.41/0.875 | 29.60/0.898 | 29.10/0.869 | 30.76/0.915 |
| Rain100L | 19.45/0.754 | 36.58/0.977 | 31.34/0.930 | 36.51/0.976 |
| Rain12 | 16.45/0.624 | 36.33/0.961 | 34.44/0.947 | 36.29/0.960 |

TABLE III. DERAINING COMPARISON OF MH-DERAINNET WITH DIFFERENT VALUES OF $\lambda_p$.

| Dataset | $\lambda_p = 0.1$ | $\lambda_p = 10$ | $\lambda_p = 1$ |
|---|---|---|---|
| Rain100H | 29.23/0.887 | 29.41/0.902 | 30.76/0.915 |
| Rain100L | 37.10/0.978 | 37.10/0.977 | 36.97/0.976 |
| Rain12 | 36.43/0.960 | 36.47/0.960 | 36.46/0.959 |

TABLE IV. DERAINING COMPARISON OF MH-DERAINNET WITH DIFFERENT NUMBERS OF STREAMS.

| Dataset | MHDN-ss | MHDN-ds | MHDN-ts | MHDN-fs |
|---|---|---|---|---|
| Rain100H | 27.82/0.880 | 29.24/0.891 | 30.76/0.915 | 30.39/0.912 |
| Rain100L | 35.90/0.973 | 36.51/0.975 | 36.64/0.976 | 36.93/0.977 |
| Rain12 | 36.08/0.959 | 36.33/0.960 | 36.31/0.959 | 36.32/0.957 |

TABLE V. DERAINING COMPARISON OF MH-DERAINNET WITH DIFFERENT NUMBERS OF CASCADED DPRDBs.

| Dataset | MHDN-4 | MHDN-5 | MHDN-6 | MHDN-7 |
|---|---|---|---|---|
| Rain100H | 29.85/0.902 | 29.79/0.901 | 30.76/0.915 | 30.56/0.913 |
| Rain100L | 36.23/0.958 | 36.25/0.962 | 36.44/0.975 | 36.33/0.971 |
| Rain12 | 35.99/0.940 | 35.98/0.943 | 36.12/0.958 | 36.12/0.952 |

*C. An Ablation Study*

In the ablation study, we will discuss the importance of different components in our MH-DerainNet architecture. Specifically, we present an ablation study on loss function, tradeoff parameters, number of streams and number of the cascaded DPRDBs on Rain100H, Rain100L and Rain12.

**1) Loss Function:** We first discuss the effectivenss of the loss function in our MH-DerainNet. In this study, we simply evaluate the result using single MSE loss, single SSIM loss, mixed MSE and perceptual loss with the ratio being 1:1 and mixed SSIM and perceptual loss with the ratio being 1:1 respectively in MH-DerainNet for performance comparison. Note that these forms of losses are denoted as $L_{mse}$, $L_{ssim}$, $L_{mse} + L_p$ and $L_{ssim} + L_p$ respectively. The deraining results (PSNR/SSIM) of MH-DerainNet with difference losses are shown in TABLE II. We can find that $L_{ssim} + L_p$ obtains much better performance on the difficult Rain100H dataset, which is the main reason why we choose the hybrid loss over SSIM loss and perceptual loss to preserve the per-pixel similarity as well as preserving the global structures. $L_{ssim}$ works well by delivering comparable results to $L_{ssim} + L_p$ on Rain100L and Rain12, and are superior to $L_{mse}$ and $L_{mse} + L_p$ for deraining.

**2) Tradeoff parameter between SSIM and perceptual losses:** In this study, we investigate the selection of suitable value of $\lambda_p$ in Eqn.(12) for trading-off the SSIM loss and perceptual loss. To compare the results, we set $\lambda_p$ to 0.1, 1 and 10 respectively. The comparison results of deraining by applying different values of $\lambda_p$ are shown in TABLE III. We can find that our MH-DerainNet with $\lambda_p = 1$ obtains the best records on Rain100H and Rain12, while MH-DerainNet with $\lambda_p = 0.1$ or $\lambda_p = 10$ cannot obtain highly-competitive results. Hence, we simply fix $\lambda_p$ to 1 in this paper.

**3) Number of streams in the multi-stream process:** We examine the deraining results of using MH-DerainNet with different numbers of streams to illustrate that concatenating multi-scaled features from multi-streams can deliver better presentation of negative rain streaks so that the background of rainy image can be acquired by addtion to removing rain streaks from images. The results of using MH-DerainNet with different numbers of streams are shown in TABLE IV, where MHDN-ss denotes MH-DerainNet that has single stream of kernel size of 3×3, MHDN-ds denotes MH-DerainNet that has two streams of kernel sizes of 3×3 and 5×5 respectively, MHDN-ts is the proposed framework having three streams with kernel sizes of 3×3, 5×5 and 7×7 respectively, and MHDN-fs denotes MH-DerainNet that has four streams of kernel sizes of 1×1, 3×3, 5×5 and 7×7 respectively. From the results, we find that MH-DerainNet with 3 streams obtains the promissing results while not costing so much GPU memory as MHDN-fs, so the number of streams is set to 3 in this study.

**4) Number of cascaded DPRDBs in multi-stream process:** We also evaluate the deraining results of MH-DerainNet with different numbers of cascaded DPRDBs. The results of under different numbers of cascaded DPRDBs are shown in TABLE V, where MHDN-i denotes our MH-DerainNet with $i$ cascaded DPRDBs in each stream. We find that the setting of 6 cascaded DPRDBs in each stream achieves best performance. With the growing numbers of the cascaded DPRDBs, the GPU memory cost will increase fast and the performance also decreases. As such, we simply set the number of cascaded DPRDBs to 6 in each stream of our framework.

## V. CONCLUDING REMARKS AND FUTURE WORK

In this paper, we have discussed the single image deraining problem and moreover have proposed a novel coarse-to-fine deraining framework called Multi-stream Hybrid Deraining Network with hybrid loss. In comparison to recent deraining methods which attempt to use residual block or dense block as an extractor, we present a new rain streaks extractor called dual path residual dense block which can obtain the negative rain streaks during training process more precisely, since the dual path block can not only reuse common features from the previous layers but also explore new features. We also use the multi-stream with shortcuts between the cascaded dual path residual dense blocks based streams, which can concatenate different scaled features to improve the deraining results. The hybrid loss can also preserve the per-pixel similarity as well as preserving the global structures, which can also enable the deraining model to obtain more distinct derained images.

Extensive experiments and comparisons are conducted on synthetic and real rainy images to evaluate MH-DerainNet. The investigated results show that our model can outperform several recent related methods. In future work, we will further explore how to reduce the parameters of our MH-DerainNet so that it can be deployed on lightweight devices.

ACKNOWLEDGMENT

This work is partially supported by the National Natural Science Foundation of China (61672365, 61732008, 61725203, 61622305, 61871444, 61572339), and Fundamental Research Funds for Central Universities of China (JZ2019HGPA01-02). Dr. Zhao Zhang is the corresponding author.